\documentclass{jfm}
\usepackage{graphicx,natbib}

\ifCUPmtlplainloaded \else
  \checkfont{eurm10}
  \iffontfound
    \IfFileExists{upmath.sty}
      {\typeout{^^JFound AMS Euler Roman fonts on the system,
                   using the 'upmath' package.^^J}%
       \usepackage{upmath}}
      {\typeout{^^JFound AMS Euler Roman fonts on the system, but you
                   dont seem to have the}%
       \typeout{'upmath' package installed. JFM.cls can take advantage
                 of these fonts,^^Jif you use 'upmath' package.^^J}%
      }
  \else
  \fi
\fi

\ifCUPmtlplainloaded \else
  \checkfont{msam10}
  \iffontfound
    \IfFileExists{amssymb.sty}
      {\typeout{^^JFound AMS Symbol fonts on the system, using the
                'amssymb' package.^^J}%
       \usepackage{amssymb}%
       \let\le=\leqslant  
         
      }{}
  \fi
\fi

\ifCUPmtlplainloaded \else
  \IfFileExists{amsbsy.sty}
    {\typeout{^^JFound the 'amsbsy' package on the system, using it.^^J}%
     \usepackage{amsbsy}}
    {}
\fi

\newsavebox{\astrutbox}
\sbox{\astrutbox}{\rule[-5pt]{0pt}{20pt}}

\title[On the aggregation of inertial particles in random flows]{On the
aggregation of inertial particles in random flows}

\author[B. Mehlig, M. Wilkinson, K. Duncan, T. Weber and M. Ljunggren]
{B.\ns M\ls E\ls H\ls L\ls I\ls G$^1$,
M.\ns W\ls I\ls L\ls K\ls I\ls N\ls S\ls O\ls N$^2$,
K.\ns  D\ls U\ls N\ls C\ls A\ls N$^2$,
T.\ns  W\ls E\ls B\ls E\ls R$^1$,
\and
M.\ns L\ls J\ls U\ls N\ls G\ls G\ls R\ls E\ls N$^1$}

\affiliation{$^1$Theoretical Physics, 
School of Physics and Engineering Physics,
G\"oteborg University/Chalmers, 41296 G\"{o}teborg, Sweden\\[\affilskip]
$^2$Department of Applied Mathematics, 
The Open University, Walton Hall, Milton Keynes, MK7 6AA, England}

\pubyear{}
\volume{}
\pagerange{}
\date{??}
\begin{document}

\maketitle

\begin{abstract}
We describe a criterion for particles suspended in a 
randomly moving fluid to aggregate. Aggregation
occurs when the expectation value of a random variable 
is negative. This variable evolves
under a stochastic differential equation.
We analyse this equation in detail in the limit where the
correlation time of the velocity field of the fluid is very short, such that
the stochastic differential equation is a Langevin equation.
\end{abstract}

\section{Introduction}
\label{sec: 1}
\par
\par
\noindent{\sl 1.1 Illustration and context}
\par
\begin{figure}
\vspace*{4mm}
\centerline{\includegraphics[width=4.15cm,clip]{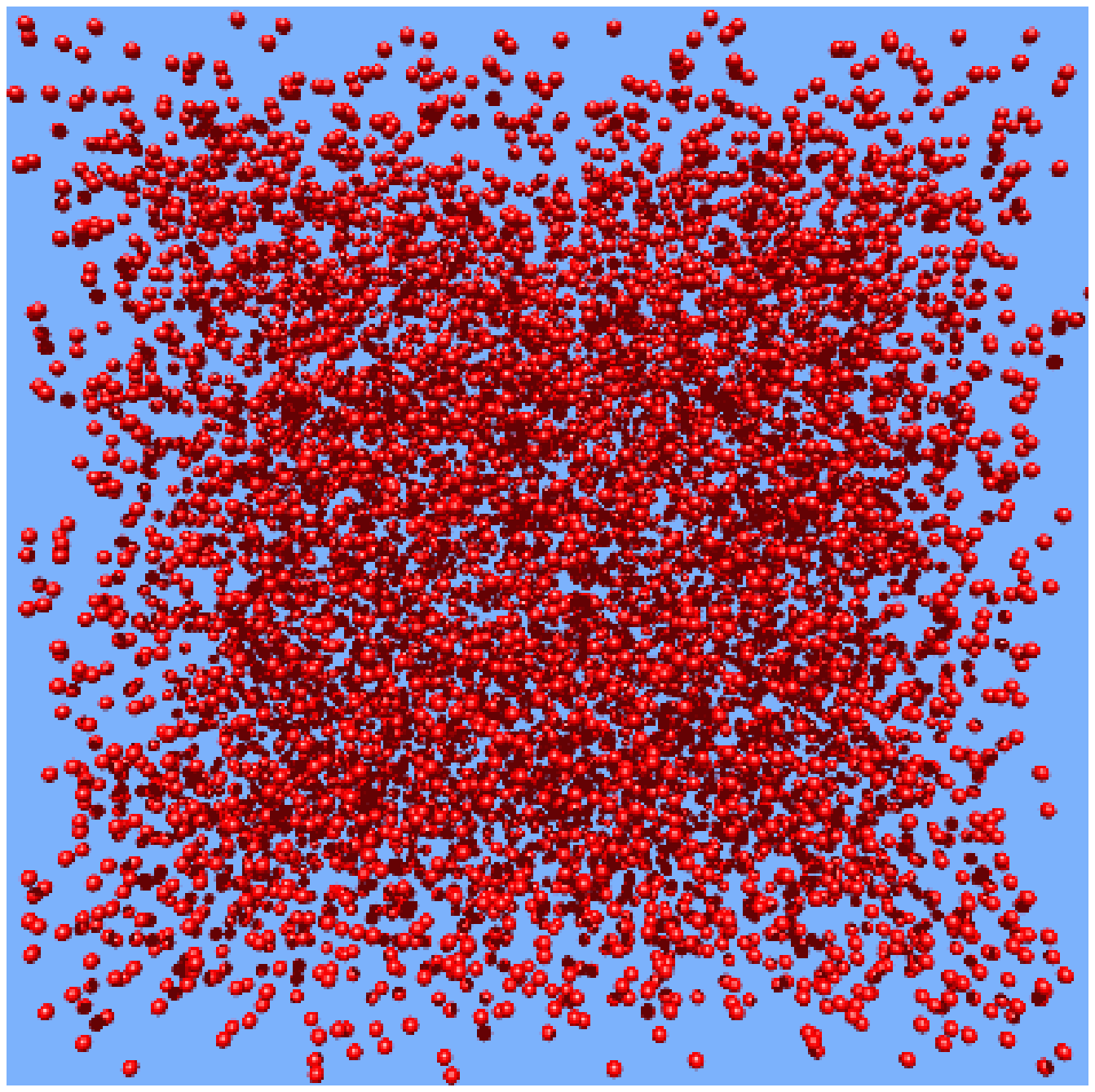}
\hspace*{4mm}\includegraphics[width=4.15cm,clip]{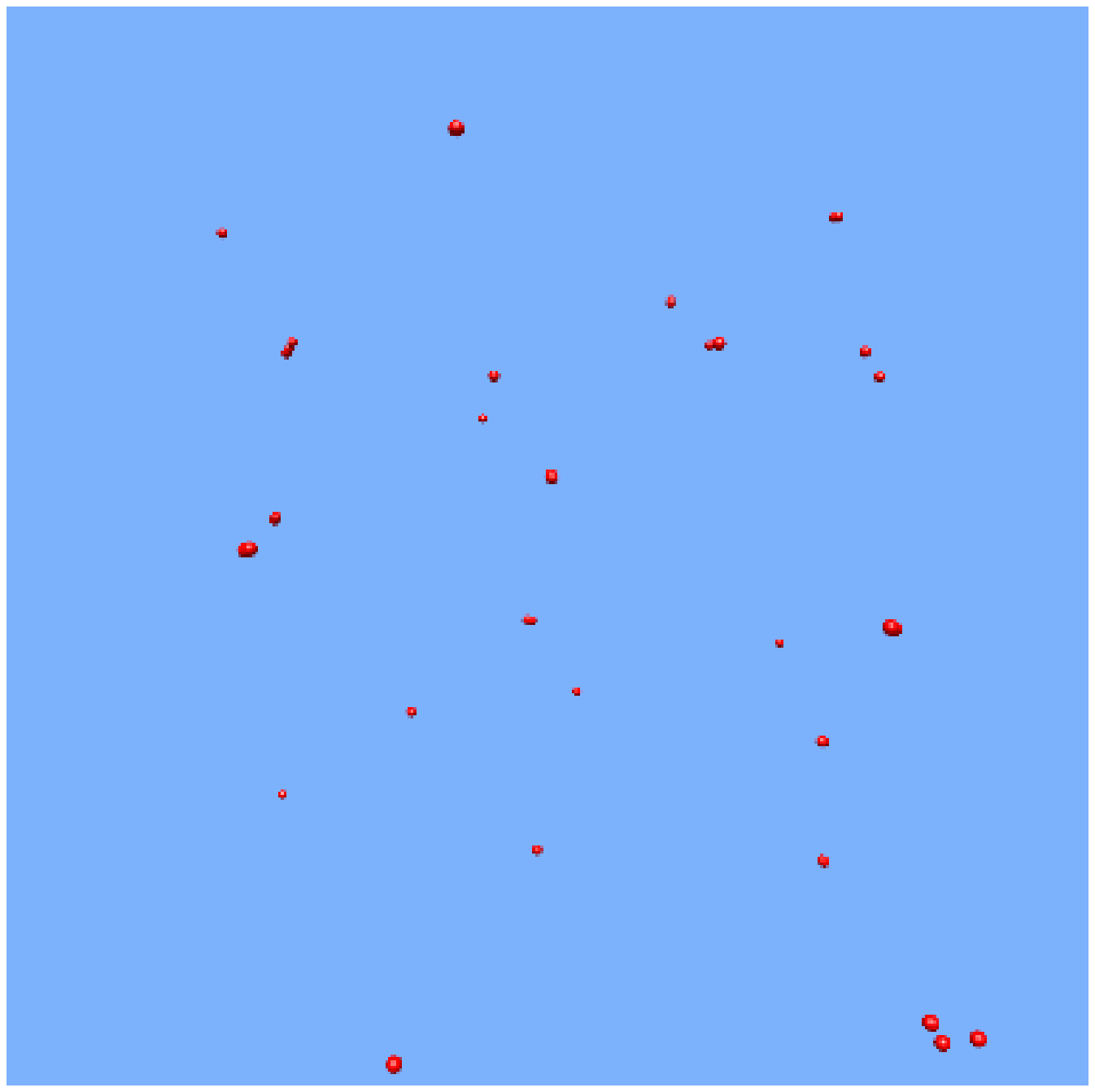}
\hspace*{4mm}\includegraphics[width=4.15cm,clip]{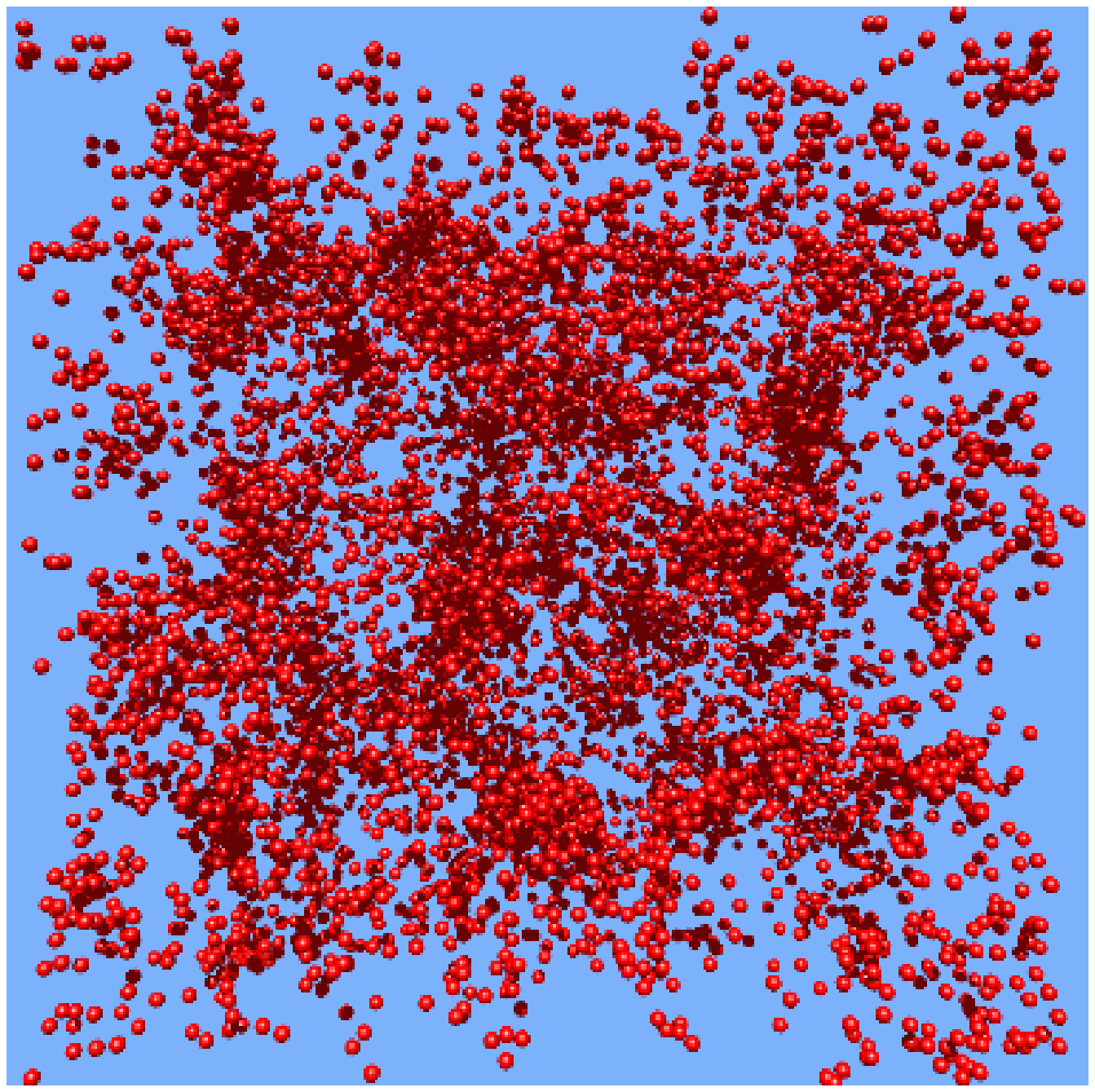}}
\caption{\label{fig: 1}
Illustrating the aggregation of non-interacting particles
in a random three-dimensional flow: the motion
is defined by equations (\ref{eq: 1.1}) to (\ref{eq: 1.3}),
(using a simplified form explained in section \ref{sec: 3}:
see equation (\ref{eq: 3.23})). The left
panel shows the initial configuration at $t=0$, the center and right panels
show final configurations (at $t=180\,\tau$) for two different
values of $\gamma$.
In the case shown in the center panel,
trajectories coalesce, until eventually all particles
follow the same trajectory. The particles in the right
panel exhibit density
fluctuations, but the trajectories do not coalesce.
}
\end{figure}

Figure~\ref{fig: 1}
illustrates a simulation of the 
distribution of small particles suspended in a three-dimensional
random flow: the particles are modelled as points, but they
are shown as small spheres to make them visible in the figure. 
The suspended particles do not interact (that is, the motion of each
particle is independent of the coordinates of the other particles, 
the equations of motion are given below).
We show the initial configuration (Figure~\ref{fig: 1}{\bf a}),
and two snapshots of the particle positions after a long time, 
with differing values of the fluid viscosity, 
(Figures~\ref{fig: 1}{\bf b} and {\bf c}).  In one case
the particles aggregate, in the sense that the trajectories of 
different particles coalesce.  In the other their distribution
shows some degree of clustering, but their trajectories never
coalesce.
In this paper we present an analysis of the transition between 
aggregating and non-aggregating phases, 
which we term the \lq path-coalescence transition'.

There are numerous experimental observations that
when small particles are suspended in a complex and apparently
random flow, their density becomes non-uniform. 
Clustering of particles into regions of high density
has been observed in experiments on particles floating on the surface
of liquids with a complex or turbulent flow \cite[]{Som93,Cre04}, and 
also turbulent flow in channels \cite[]{Fes94,vHa98}. 
The conditions 
under which this occurs are not yet fully understood. 
The aggregation effect illustrated in figure 1 is 
an extreme form of clustering. 
There appears to be less experimental work on 
aggregation, but coalescence of suspended water droplets 
is clearly very important in the formation of rain drops 
in clouds \cite[]{Sha03}.
%This process is still not very well understood, 
Even less is known about aggregation. In particular
it is not clear when a model of particles suspended in a random flow
can exhibit aggregation, and when additional physical
phenomena (such as differential drift velocities under gravitational
forces, or Brownian diffusion) must be invoked.

Earlier theoretical work on clustering in random flows
has used Fokker-Planck equations determining 
moments of the particle density of advected
particles \cite[]{Kly99}. This \lq passive scalar' 
approach does not allow for the effects of
inertia of the particles. It has been supplemented by a 
perturbative analysis of the deviations of particles from 
advected trajectories, as proposed by \cite{Max87}.
These two approaches are combined in papers by 
\cite{Elp96} and \cite{Fal03}. 
Numerical work indicates that clustering
in solenoidal flows occurs most readily when the correlation
time of the velocity field is comparable with the 
time constant associated with the viscous drag \cite[]{Sig02}.
The aggregation (as opposed 
to clustering) of
particles by random flows has received relatively little
attention. \cite{Deu85} appears to
have been the first to propose that particles
subjected to a smooth random flow can coalesce, and showed
numerical evidence that this can happen in one dimension.
He argued that there is a transition between coalescing 
and non-coalescing phases, and identified the dimensionless 
parameter which determines the phase transition in one dimension.

This paper describes results obtained from a new approach
to characterising particle aggregation in random flows. 
It is based upon calculating a Lyapunov exponent describing
the rate of separation of nearby particles from a solution 
of a system of stochastic differential equations: the Lyapunov 
exponent is the expectation value of one of the variables.
In the limit as the correlation time of the flow approaches
zero, the stochastic differential equations can be reduced
to a pair of Langevin equations. 
Our results 
for three-dimensional flows which are discussed here build upon earlier work 
by two of the present authors in one dimension 
\cite[]{Wil03} (where we solved Deutsch's model
exactly) and two dimensions \cite[]{Meh04}.
The three-dimensional case is most important
for physical applications, but involves substantial additional
technical complications.

We remark that \cite{Pit01} also considered the two-dimensional 
case, and quotes an analytic expression for the maximal
Lyapunov exponent. His expression is
incorrect in two dimensions, because the generating function
that he uses is divergent for the equilibrium distribution.
His calculation can be adapted to give the correct expression 
in the one dimensional case, as quoted in \cite{Meh04}.

The definition of Lyapunov exponents is explained by 
\cite{Eck79}, who also discuss the method we use for
extracting the Lyapunov exponents from direct numerical simulations. 
We note that in the case where inertia of the particles can be neglected, 
our results reduce to calculating the Lyapunov exponent for a spatially
correlated Brownian motion, which was discussed
from a mathematical point of view by \cite{LeJ85}
and \cite{Bax86}. 
\par
\ 
\par
\
\par
\noindent{\sl 1.2 The model}
\par
The most natural model for theoretical investigation
is the motion of spherical particles (radius $a$, mass $m$) 
moving in a random velocity field ${\bf u}({\bf r},t)$ with 
specified isotropic and homogeneous statistical properties. 
The particles are assumed not to affect either the flow,
or each other's motion, and to experience a drag force
given by Stokes's law: the force ${\bf f}_{\rm dr}$ on a 
particle moving with velocity ${\bf v}$ relative to the 
fluid is ${\bf f}_{\rm dr}=-6\pi \eta a{\bf v}$, where $\eta$
is the viscosity of the fluid.
We will simplify the problem by assuming that the 
particles are made of a material which is much
denser than the fluid in which they are suspended:
this enables us to neglect the inertia of the displaced
fluid. Accordingly, we consider a large number of suspended
particles, initially with random positions and zero
velocity, having equations of motion
\begin{eqnarray}
\label{eq: 1.1}
\dot{\bf r}&=&{1\over{m}}{\bf p}\ ,\qquad
\dot {\bf p}=-\gamma [{\bf p}-m{\bf u}({\bf r},t)]\,.
\end{eqnarray}
The random velocity field ${\bf u}({\bf r},t)$ could be either
externally imposed (for example, if a gas is driven by
an ultrasonic noise source), or self-generated (as in the
case of turbulence). The equations of motion with displaced
mass effects included are discussed by \cite{Lan58}.
Our neglect of displaced mass effects is justified for
aerosol systems.

In order to fully specify the problem we must define
the statistical properties of the random velocity field
${\bf u}({\bf r},t)$.
The random force ${\bf f}=m\gamma{\bf u}$ is generated 
from a vector potential ${\bf A}=(A_1,A_2,A_3)$ and a scalar 
potential $\phi=A_0$:
\begin{equation}
\label{eq: 1.2}
{\bf f}=\nabla \phi+\nabla \wedge {\bf A}
\ .
\end{equation}
The scalar fields $A_i({\bf r},t)$ have isotropic, homogeneous and
stationary statistics. We assume that these fields are 
statistically independent, and that they 
all have the same correlation function,
except that the intensity of $\phi=A_0$ exceeds that of the 
other fields by a factor $1/\alpha^2$:
\begin{equation}
\label{eq: 1.3}
\langle A_i({\bf r},t)A_j({\bf r}',t')\rangle=
\delta_{ij}[(1-\alpha^2)\delta_{i0}+\alpha^2]
C(\vert{\bf r}-{\bf r}'\vert,t-t')
\ .
\end{equation}
The random force ${\bf f}({\bf r},t)$ is characterised by its typical
magnitude $\sigma$, and by the correlation length $\xi$ and correlation
time $\tau$ of the correlation function $C(R,t)$. In the case 
of a well-developed turbulent flow, the velocity field
has a power-law spectrum with upper and lower cutoffs
\cite[]{Fri97}.
If a random velocity field modeling fully-developed 
turbulence is used in our model, it is most appropriate
to take the correlation length and time to be those of 
the \lq dissipation scale', that is, the cutoff with the 
smaller length scale.

The system of equations is
characterised by three independent dimensionless parameters,
which we take as
\begin{equation}
\label{eq: 1.4}
\nu=\gamma \tau\ ,\ \  \chi={\sigma \tau^2\over{m\xi}}\ ,\ \ \ \alpha
\ .
\end{equation}
The parameter $\nu $ is a dimensionless measure of the degree of
damping, 
$\chi $ is a dimensionless measure of the strength of the forcing
term and $\alpha $ measures the relative magnitudes of potential
and solenoidal components of the velocity field (which is purely
potential when $\alpha=0$, and purely solenoidal in the limit as
$\alpha \to \infty$).  
\par
\ 
\par
\ 
\par
\
\par
\noindent{\sl 1.3 Description of our results}
\par
We show that the phase transition is determined by a
Lyapunov exponent, $\lambda_1$, describing the separation
of nearby particles: their trajectories coalesce if $\lambda_1<0$. 
Here we describe a new and general approach 
to this problem, reducing the determination of the
Lyapunov exponent to the analysis of a simple 
dynamical system, described by a system of 
ordinary differential equations containing stochastic
forcing terms: the Lyapunov exponent is found to be proportional
to the expectation value of one coordinate.
These stochastic differential 
equations are derived in section \ref{sec: 2}. They
introduce an apparent paradox: the structure of the equations
appears to be identical in  two-dimensional
and three-dimensional flows, suggesting that 
the path-coalescence transition is fundamentally
the same in two and three dimensions. That would be
a very surprising conclusion.

In order to demonstrate and illuminate our method, in sections 
\ref{sec: 3} and \ref{sec: 4} 
we  pursue the solution of this problem in considerable
depth in one limiting case, namely the limit where the correlation
time $\tau $ of the random velocity field is small and the random
force is sufficiently weak 
(strictly, we consider the case where $\chi \ll \nu \ll 1$).
In this limit, the system of ordinary
differential equations described in section \ref{sec: 2} becomes a
system of two coupled Langevin equations. 
In section \ref{sec: 3} we show that there is in fact a 
difference between the three-dimensional problem and the 
two-dimensional case studied in \cite{Meh04}, which is a rather subtle
example of the difficulties in applying Langevin approaches
to nonlinear equations \cite[]{vKa92}.

In section \ref{sec: 4} we discuss a perturbation theory for the 
Lyapunov exponent describing the phase transition, expanded
in powers of a parameter 
$\epsilon=\chi\nu^{-3/2}$ which is a dimensionless
measure of the inertia of the particles. The perturbation
theory is constructed by transforming the Langevin equation
first into a Fokker-Planck equation, and then into a
non-Hermite an perturbation of a three-dimensional isotropic
quantum harmonic oscillator. We are then able to use the 
harmonic-oscillator creation and annihilation operators 
\cite[]{Dir30} to 
express the perturbation theory in a purely algebraic
form, enabling us to compute the coefficients to any desired
order. We investigate the phase diagram (the line in parameter space
separating coalescing and non-coalescing phases), using
both Monte Carlo averaging of the Langevin equation 
and the results of our 
perturbation theory. We find that aggregation can only
occur if the random flow has a certain degree of compressibility, 
which increases as the effects of inertia increase, until
there is no coalescing phase even for a purely potential flow field.

The analysis in sections \ref{sec: 2} to \ref{sec: 4} 
considers the case where
all of the suspended particles have the same mass $m$ and damping
rate $\gamma$.
This ideal can be approached quite accurately in model experiments,
but in most applications in the natural world and in technology,
the suspended particles will have different masses, sizes and shapes.
In section \ref{sec: 5} we discuss the effect of dispersion in the 
distribution of masses in the one-dimensional case: these 
arguments can be adapted to treat higher dimensions and
dispersion of the damping constant $\gamma$. 
We argue that 
path coalescence is not destroyed by mass dispersion
(although of course it is no longer a sharp transition). 
In this paper we give a quite comprehensive discussion of the
the case where the correlation time of the random flow
is very short, and the stochastic differential equations
derived in section 2 can be approximated by Langevin equations.
Section 6 discusses how our approach can be extended to
other cases.

\section{Equations determining the Lyapunov exponent}
\label{sec: 2}

To determine whether particles cluster together, 
we consider two nearby trajectories with spatial
separation $\delta {\bf r}$ and momenta differing by $\delta {\bf p}$. The
linearised equations of motion derived from (\ref{eq: 1.1}) are
\begin{eqnarray}
\label{eq: 2.1}
\delta \dot {\bf r}&=&{1\over m}\delta {\bf p}\,,\qquad
\delta \dot {\bf p}=-\gamma \delta {\bf p}+\tilde F(t)\delta {\bf r}
\ .
\end{eqnarray}
Here $\tilde F(t)$ is proportional to the strain-rate matrix
of the velocity field, with elements
\begin{equation}
\label{eq: 2.1a}
F_{ij}(t)={\partial f_i\over{\partial r_j}}({\bf r}(t),t)=m\gamma
{\partial u_i\over{\partial r_j}}({\bf r}(t),t)
\ .
\end{equation}
It is convenient to
parameterise $\delta {\bf r}$ and $\delta {\bf p}$ as follows:
\begin{eqnarray}
\label{eq: 2.2}
\delta {\bf r}&=&X{\bf n}_1\,,\qquad
\delta {\bf p}=X(Y_1{\bf n}_1+Y_2{\bf n}_2)\,,
\end{eqnarray}
where ${\bf n}_1$ and ${\bf n}_2$ are orthogonal unit vectors, which depend
upon time. 
The parameter $X$ is a scale factor: trajectories coalesce if 
$X$ decreases with probability unity in the long-time limit.
In the three-dimensional case, we  find it convenient 
to introduce the third
element 
${\bf n}_3={\bf n}_1\wedge {\bf n}_2$ of a time-dependent orthonormal basis 
so that 
\begin{eqnarray}
\label{eq: 2.4}
{\bf n}_i.{\bf n}_j&=&\delta_{ij}
\qquad \mbox{and}\qquad
{\bf n}_i=\varepsilon_{ijk}{\bf n}_j\wedge{\bf n}_k
\ .
\end{eqnarray}
Differentiating (\ref{eq: 2.2}), and substituting the resulting 
expressions into (\ref{eq: 2.1}) gives
\begin{eqnarray}
\label{eq: 2.5}
\delta \dot {\bf r}&=&\dot X{\bf n}_1+X \dot {\bf n}_1
\nonumber \\
&=&{1\over m}X(Y_1{\bf n}_1+Y_2{\bf n}_2)
\\
\delta \dot {\bf p}&=&\dot X(Y_1{\bf n}_1+Y_2{\bf n}_2)
+X(\dot Y_1{\bf n}_1+\dot Y_2{\bf n}_2)+X(Y_1\dot{\bf n}_1+Y_2\dot {\bf n}_2)
\nonumber \\
&=&-\gamma X(Y_1{\bf n_1}+Y_2{\bf n_2})+\tilde F(t){\bf n}_1
\ .
\nonumber
\end{eqnarray}
Projecting $\delta \dot {\bf r}$ onto the unit vectors ${\bf n}_i$
gives the following three scalar equations of motion
\begin{eqnarray}
{\bf n}_1.\delta \dot {\bf r}&=&\dot X={1\over m}Y_1 X
\nonumber
\\
\label{eq: 2.6}
{\bf n}_2.\delta \dot {\bf r}&=&X({\bf n}_2.\dot {\bf n}_1)={1\over m}Y_2X
\\
{\bf n}_3.\delta \dot {\bf r}&=&X \dot {\bf n}_1.{\bf n}_3=0
\ .
\nonumber
\end{eqnarray}
The last of these equations implies that $\dot {\bf n}_1\wedge{\bf n}_2=0$,
so that $\dot {\bf n}_1$ is proportional to ${\bf n}_2$: we write 
\begin{equation}
\label{eq: 2.7}
\dot {\bf n}_1=\dot \theta {\bf n}_2
\end{equation}
so that the equation for ${\bf n}_2.\delta \dot {\bf r}$ gives
\begin{equation}
\label{eq: 2.8}
\dot \theta={1\over m}Y_2
\ .
\end{equation}
The first equation of (\ref{eq: 2.6}) indicates that $X$ is a 
product of random variables, and therefore has a log-normal distribution,
that is, the logarithm of $X$ has a Gaussian probability density.
In the limit as $t\to \infty$, the mean and variance of $\log_{\rm e}X$
are both linear functions of time:
\begin{equation}
\label{eq: 2.9a}
\langle \log_{\rm e}X\rangle\sim \lambda_1 t+c_1\ ,\ \ \ 
{\rm var}(X)=\mu t+c_2
\end{equation}
where $\lambda_1$, $\mu$, $c_1$ and $c_2$ are constants.
If $\lambda_1<0$, the probability of $\log_{\rm e}X$ exceeding any
specified value approaches zero as $t\to \infty$, implying that
trajectories of nearby particles almost always coalesce.

The Lyapunov exponent $\lambda_1$ is the mean value of the derivative
${\rm d}\log_{\rm e}X/{\rm d}t$, so that the first equation
of (\ref{eq: 2.6}) gives
\begin{equation}
\label{eq: 2.9}
\lambda_1={1\over m}\langle Y_1\rangle
\ .
\end{equation}
Now consider the three projections of $\delta \dot {\bf p}$, as given
by equation (\ref{eq: 2.5}):
\begin{eqnarray}
{\bf n}_1.\delta \dot {\bf p}&=&\dot XY_1+X\dot Y_1
+XY_2({\bf n}_1.\dot {\bf n}_2)
\nonumber \\
&=&-\gamma XY_1+{\bf n}_1.\tilde F(t){\bf n}_1X
\nonumber
\\
\label{eq: 2.10}
{\bf n}_2.\delta \dot {\bf p}&=&\dot XY_2+X\dot Y_2
+XY_1(\dot {\bf n}_1.{\bf n}_2)
\nonumber \\
&=&-\gamma XY_2+{\bf n}_2.\tilde F(t){\bf n}_1X
\\
{\bf n}_3.\delta \dot {\bf p}&=&
XY_1({\bf n}_3.\dot {\bf n}_1)+XY_2({\bf n}_3.\dot {\bf n}_2)
\nonumber \\
&=&{\bf n}_3.\tilde F(t){\bf n}_1 X
\nonumber
\ .
\end{eqnarray}
We introduce the notation 
\begin{equation}
\label{eq: 2.11}
{\bf n}_i(t).\tilde F(t){\bf n}_j(t)=F'_{ij}(t)
\end{equation}
and note that the statistics of the transformed matrix elements
$F'_{ij}(t)$ are the same as those of the original elements
$F_{ij}(t)$, because the statistics of the velocity field
are isotropic.
Using eqs.~(\ref{eq: 2.6}) to (\ref{eq: 2.8}) and 
$({\bf n}_i.\dot {\bf n}_j)+(\dot {\bf n}_i.{\bf n}_j)=0$ to simplify, we
find the following equations of motion for the variables $Y_i$
\begin{eqnarray}
\label{eq: 2.12}
\dot Y_1&=&-\gamma Y_1+{1\over m}(Y_2^2-Y_1^2)+F'_{11}(t)
\\
\dot Y_2&=&-\gamma Y_2-{2\over m}Y_1Y_2+F'_{21}(t)
\nonumber
\ .
\end{eqnarray}
Finally, the equation for ${\bf n}_3.\delta \dot {\bf p}$ gives
\begin{equation}
\label{eq: 2.13}
{\bf n}_2.\dot {\bf n}_3=-{1\over Y_2}F'_{31}(t) 
\ .
\end{equation}
Eqs.~(\ref{eq: 2.9}) and (\ref{eq: 2.12}) are the principal results
of this paper. Eq.~(\ref{eq: 2.9}) shows that the Lyapunov exponent
(the sign of which determines whether or not path coalescence occurs) 
is given      
by the expectation value of a random variable $Y_1$ of a simple,
finite dimensional stochastic dynamical system, described
by eqs.~(\ref{eq: 2.12}). This dynamical system is 
almost completely de-coupled from the other variables:
the equations for $Y_1$ and $Y_2$ do not depend upon $X$, and the
vectors ${\bf n}_i(t)$ only enter these equations through the evaluation
of the random matrix elements $F'_{ij}$. We pointed out that 
statistics of these elements are independent of the orientation
of the orthogonal triplet $({\bf n}_1,{\bf n}_2,{\bf n}_3)$.

In the two-dimensional case the analysis leading to (\ref{eq: 2.12}) 
proceeds along similar lines, and leads to the same
pair of equations \cite[]{Meh04}. 
The only difference 
is that the equation (\ref{eq: 2.13}) is absent
in the two-dimensional case.
This suggests that the expression
for the Lyapunov exponent $\lambda_1=\langle Y_1\rangle/m$ 
should be the same in two and three dimensions.
This would be a surprising conclusion, but it is not obvious how 
it can be averted. However, it does prove to be false,
as will be demonstrated in the next section for the limiting case
where
%$\nu\ll 1$ and $\chi/\nu \ll 1$.
$\chi \ll \nu \ll 1$.

\section{The Langevin approximation}
\label{sec: 3}

Let us now consider how to treat
eqs.~(\ref{eq: 2.12}) in the limit where the correlation
time $\tau$ of $\tilde F(t)$ is very short.
%(that is, $\nu \ll 1$). 
Because the random field ${\bf f}({\bf r},t)$ is fluctuating very rapidly, 
the position ${\bf r}(t)$ of a particle at time $t$ is independent
of the instantaneous value of the force ${\bf f}({\bf r},t)$, 
so that the value
if $F_{ij}(t)=\partial f_i/\partial x_j({\bf r}(t),t)$ at the position
of the particle is statistically indistinguishable from a random
sample of the field $\partial f_i/\partial x_j$.
We also assume that the gradients of the fluctuating forces
(the quantities $F'_{11}$ and $F'_{22}$ in equations (\ref{eq: 2.12}))
are sufficiently small that the typical magnitude of the 
displacement of the variables $Y_1$, $Y_2$ occurring during 
the correlation time $\tau$ is small compared to the typical
magnitude of these variables. This condition is expressed
in terms of the dimensionless variables $\chi$ and $\nu$
later in this section. Under these conditions of short
correlation time and small amplitude, 
equations (\ref{eq: 2.12}) may be replaced by 
Langevin equations. At first sight, this would appear to lead
to
\begin{eqnarray}
\label{eq: 3.1}
{\rm d}Y_1&=&\biggl[-\gamma Y_1+{1\over m}(Y_2^2-Y_1^2)\biggr]{\rm d}t
+{\rm d}f_1
\\
{\rm d}Y_2&=&\biggl[-\gamma Y_2-{2\over m}Y_1Y_2\biggr]{\rm d}t+{\rm d}f_2
\nonumber
\end{eqnarray} 
where the ${\rm d}f_i$ are increments of a Brownian process, satisfying
$\langle {\rm d}f_i\rangle=0$ and 
$\langle {\rm d}f_i{\rm d}f_j\rangle=2D_{ij}{\rm d}t$, for
some constant diffusion coefficients $D_{ij}$.
In the two-dimensional case, this expectation is correct
\cite[]{Meh04}, and
eqs.~(\ref{eq: 3.1}) are the appropriate Langevin equations.
In the three-dimensional case, we will see that an additional
drift term must be added to the second of these equations.
This is a consequence of the fact that, in the three-dimensional
case, $Y_2$ is a non-linear function of the components of the vector
$\delta {\bf p}-Y_1\delta {\bf r}$ because it is the magnitude
of this vector. This is an example of the difficulties that
arise when treating Langevin equations involving non-linear 
functions of noise terms \cite[]{vKa92}.

In order to determine the correct Langevin equations to
model (\ref{eq: 2.12}), 
let us consider the integral of the stochastic forcing terms
${\rm d}f_i$ over a time
interval $\delta t$ which is long compared to $\tau$, but
short enough that the change in the variables $Y_i$ occurring
in time $\delta t$ can be neglected. We define
\begin{equation}
\label{eq: 3.2}
\delta f_i(t)=\int_t^{t+\delta t}\!\!\!\!{\rm d}t'\ F'_{i1}(t')
\end{equation}
and find
\begin{equation}
\label{eq: 3.3}
\langle \delta f_i\delta f_j\rangle=2D_{ij}\delta t+O(\tau)
\end{equation}
where
\begin{equation}
\label{eq: 3.4}
D_{ij}={\textstyle{1\over 2}}\int_{-\infty}^\infty {\rm d}t\ 
\langle F'_{i1}(t)F'_{j1}(0)\rangle
\ .
\end{equation}
The calculation of $\langle \delta f_i(t)\rangle$ is more subtle.
We have
\begin{equation}
\label{eq: 3.5}
\langle \delta f_i\rangle=\int_0^{\delta t}{\rm d}t\ 
\langle {\bf n}_i(t).\tilde F(t) {\bf n}_1(t)\rangle
\ .
\end{equation}
We must take account of the fact that the unit vectors ${\bf n}_i(t)$
are rotating: we write 
\begin{equation}
\label{eq: 3.6}
{\bf n}_i(t)=\sum_{k=1}^3 R_{ik}(t){\bf n}_k(0)
\end{equation}
where $R_{ik}(t)$ are elements of a rotation matrix.
We now write (\ref{eq: 3.5}) in the form
\begin{eqnarray}
\label{eq: 3.7}
\langle \delta f_i\rangle
&=&\int_0^{\delta t}{\rm d}t\ \sum_{k=1}^3\sum_{l=1}^3 
\langle R_{ik}(t)R_{1l}(t)F'_{kl}(t)\rangle
\ .
\end{eqnarray}
Now consider the rotation of the unit vectors: 
using (\ref{eq: 2.9}) and (\ref{eq: 2.13}) we have
\begin{eqnarray}
\nonumber      
{\bf n}_1(t)&=&{\bf n}_1(0)+\dot \theta {\bf n}_2(0)t+O(t^2)\,,
\\
\label{eq: 3.8}
{\bf n}_2(t)&=&{\bf n}_2(0)-\dot \theta {\bf n}_1(0)t+
{1\over{Y_2}}\int_0^t{\rm d}t'\ F'_{31}(t'){\bf n}_3(0)+O(t^2)\,,
\\
{\bf n}_3(t)&=&{\bf n}_3(0)
-{1\over{Y_2}}\int_0^t{\rm d}t'\ F'_{31}(t'){\bf n}_2(0)\,,
+O(t^2)
\nonumber
\end{eqnarray}
so that
\begin{equation}
\label{eq: 3.9}
\tilde R(t)=\left( \begin{array}{ccc}
1 & \dot \theta t& 0 \\
-\dot \theta t& 1 & Y_2^{-1}\int_0^t {\rm}dt'\, F'_{31} \\ 
0 & -Y_2^{-1}\int_0^t{\rm d}t'\,F'_{31}&1
\end{array} \right)
+O(t^2)
\ .
\end{equation}
We obtain 
\begin{eqnarray}
\label{eq: 3.10}
\langle \delta f_i\rangle&=&\int_0^{\delta t}{\rm d}t'\ \sum_k
\langle R_{ik}(t')F'_{k1}(t')+\dot \theta t F'_{k2}(t')\rangle
\nonumber \\
&=&\int_0^{\delta t}{\rm d}t\sum_k\langle R_{ik}(t)F'_{k1}(t)\rangle
+O(\delta t^2)
\ .
\end{eqnarray}
This yields
\begin{eqnarray}
\label{eq: 3.11}
\langle \delta f_1\rangle&=&\int_0^{\delta t}{\rm d}t\
\langle F'_{11}(t)+\dot \theta t F'_{21}(t)\rangle=0
\nonumber \\
\langle \delta f_2\rangle&=&\int_0^{\delta t}{\rm d}t\
\langle -\dot \theta F'_{11}(t)+F'_{21}(t)+{1\over{Y_2}}\int_0^t{\rm d}t'\
F'_{31}(t)F'_{31}(t')\rangle
\nonumber \\
&=&{1\over{Y_2}}\int_0^{\delta t}{\rm d}t\int_0^t{\rm d}t'
\langle F'_{31}(t)F'_{31}(t')\rangle
\nonumber \\
&=&{1\over{Y_2}}D_{31}\delta t
\ .
\end{eqnarray}
The Langevin equations therefore contain an additional drift term 
due to the fact that $\langle \delta f_2\rangle$ is non-zero:
the correct Langevin equations in three dimensions are
\begin{eqnarray}
\label{eq: 3.13}
{\rm d}Y_1&=&\biggl[-\gamma Y_1+{1\over m}(Y_2^2-Y_1^2)\biggr]{\rm d}t
+{\rm d}\zeta_1\,,
\nonumber
\\
{\rm d}Y_2&=&\biggl[-\gamma Y_2+{D_{31}\over{Y_2}}-{2\over{m}}Y_1Y_2\biggr]
{\rm d}t+{\rm d}\zeta_2\ ,
\end{eqnarray}
with
\begin{eqnarray}
\label{eq: 3.14}
\langle {\rm d}\zeta_i\rangle&=&0\ ,\qquad
\langle {\rm d}\zeta_i{\rm d}\zeta_j\rangle=2D_{ij}{\rm d}t\ .
\end{eqnarray}
The diffusion constants $D_{ij}$ were defined in equation
(\ref{eq: 3.4}). Note that in two dimensions, however, (\ref{eq: 3.1})
remains valid because the term arising from the rotation
of ${\bf n}_3$ is absent.

Now consider the evaluation of the diffusion constants
in terms of the statistics of the force ${\bf f}({\bf r},t)$.
The elements of the force-gradient matrix $\tilde F$ are
\begin{equation}
\label{eq: 3.15}
F_{ij}={\partial^2\phi\over{\partial x_i\partial x_j}}+
\epsilon_{ilk}{\partial^2 A_k\over{\partial x_j\partial x_l}}
\end{equation}
where $\epsilon_{ilk}$ is the ``Kronecker $\epsilon$-symbol'' describing
the parity of the permutation of the indices $ilk$.  Define
\begin{equation}
\label{eq: 3.16}
D_0={\textstyle{1\over 2}}\int_{-\infty}^\infty {\rm d}t\ 
\biggl\langle {\partial^2\phi\over{\partial x^2}}(t)
{\partial^2 \phi\over{\partial x^2}}(0)\biggr\rangle
\ .
\end{equation}
Then define $D_1=D_{11}$, $D_2=D_{21}=D_{31}$, so that
\begin{eqnarray}
\label{eq: 3.17}
D_1&=&D_0\biggl(1+{2\alpha^2\over 3}\biggr)
\qquad\mbox{and}\qquad
D_2=D_0\biggl({1\over 3}+{4\alpha^2\over 3}\biggr)
\ .
\end{eqnarray}
We introduce a more convenient dimensionless measure of the relative 
importance of solenoidal and potential fields
\begin{equation}
\label{eq: 3.18}
\Gamma\equiv {D_2\over {D_1}}={1+4\alpha^2\over{3+2\alpha^2}}
\end{equation}
and find ${1\over 3}\le \Gamma \le 2$ in the three-dimensional
case because $0\le \alpha\le \infty$. 
It is convenient to re-scale the Langevin equations into dimensionless
form: write
\begin{equation}
\label{eq: 3.19}
{\rm d}t' =\gamma {\rm d}t
\ ,\ \ \ 
x_i=\sqrt{\gamma\over{D_i}}Y_i
\ ,\ \ \ 
{\rm d}w_i=\sqrt{\gamma\over{D_i}}{\rm d}\zeta_i
\end{equation}
and define
\begin{equation}
\label{eq: 3.19a}
\epsilon={D_1^{1/2}\over{m\gamma^{3/2}}}
\ .
\end{equation}
With these changes of variables, the Langevin equations become
\begin{eqnarray}
\label{eq: 3.20}
{\rm d}x_1&=&[-x_1+\epsilon(\Gamma x_2^2-x_1^2)]{\rm d}t'+{\rm d}w_1
\\
{\rm d}x_2&=&[-x_2+x_2^{-1}-2\epsilon x_1x_2]{\rm d}t'+{\rm d}w_2
\nonumber
\end{eqnarray}
with
\begin{eqnarray}
\label{eq: 3.21}
\langle {\rm d}w_i\rangle&=&0\,,
\qquad
\langle {\rm d}w_i{\rm d}w_j\rangle=2\delta_{ij}{\rm d}t'
\ .
\end{eqnarray}
Eqs.~(\ref{eq: 3.20},\ref{eq: 3.21}) must be solved to determine the expectation
value of $x_1$ in the steady state. The Lyapunov exponent is then given by
\begin{equation}
\label{eq: 3.22}
\lambda_1=\gamma \epsilon \langle x_1 \rangle
\ . 
\end{equation}
\par
Figure~\ref{fig: 2}{\bf a} compares the Lyapunov exponent
obtained from a Monte Carlo simulation of equations
(\ref{eq: 3.20}--\ref{eq: 3.22}) with 
a direct numerical simulation of a random flow
described by equation (\ref{eq: 1.1}).
The Lyapunov exponents 
determined from eqs.~(\ref{eq: 3.20}) and (\ref{eq: 3.21}) 
for $\Gamma = 1/3,1$ and $2$ are plotted as red lines. 
The results
are compared to numerical simulations of (\ref{eq: 1.1}), 
using a method described
in \cite{Eck79} to determine the Lyapunov exponent. 
Because we are concerned with the limit where the
correlation time $\tau$ is taken to zero, the random 
flow was generated using a discrete series
of uncorrelated random impulses, acting over
a small time step $\delta t \gg \tau$: the impulse
\begin{equation}
\label{eq: 3.23}
{\bf f}_n({\bf r}) =\int_{n\delta t}^{(n\!+\!1)\delta t}\!\!
{\rm d}t'\, {\bf f}({\bf r}_{t'},t')
\end{equation}
at time $n\,\delta t$ is taken to
be of the form (\ref{eq: 1.2}) in terms of scalar fields
$\phi_n({\bf r})$ and ${\bf A}_n({\bf r})$ satisfying
\begin{equation}
\label{eq: 3.24}
\langle \phi_n({\bf r})\phi_{n'}({\bf r}')\rangle
=\sigma^2\, \xi^2\, \delta
t\,\exp(|{\bf r}-{\bf r}'|^2/2\xi^2)\delta_{nn'}
\end{equation}
and similarly for ${\bf A}_n({\bf r})$. This implies
$D_0 = 3\sigma^2/(2m^2\gamma^3\xi^2)$.

Now we discuss the conditions under which the Langevin
equations (\ref{eq: 3.20}) and (\ref{eq: 3.21}) are a valid 
approximation of (\ref{eq: 2.12}) and (\ref{eq: 2.13}).
For this purpose it is sufficient to consider the one-dimensional
version of equations (\ref{eq: 3.13}), namely
\begin{equation}
\label{eq: 3.25}
\dot Y=-\gamma Y-{1\over m}Y^2+F(t)
\end{equation}
(this equation appears with a different notation in \cite{Wil03}).
The Langevin equations are valid provided the changes
in the value of $Y$ over the correlation
time $\tau$ is small compared to the typical values of
this quantity. This criterion can obviously only be
satisfied if the correlation time is sufficiently short
that $\nu =\gamma\tau\ll 1$. The criterion also requires 
the stochastic force $F(t)$ to be sufficiently weak. 
The deterministic part of the velocity, $-\gamma Y-Y^2/m$, is
positive in the interval from $Y=-\gamma m$ to $Y=0$.
The criterion on the strength of $F\sim \sigma/\xi$ is
that the displacement over time $\tau$ should be small compared 
to the width of that interval, that is $\vert F\vert \tau\ll \gamma m$.
Using the fact that $\vert F\vert\sim \sigma/\xi$, we obtain
the following criteria for the validity of the Langevin
approximation:
\begin{equation}
\label{eq: 3.26}
{\chi\over {\nu }}\ll 1 \ ,\ \ \ \nu\ll 1.
\end{equation}
\par
For completeness, we end this section by mentioning how 
equations (\ref{eq: 3.20}) and (\ref{eq: 3.21}) differ in one
and -two dimensions. The one-dimensional case was considered
in \cite[]{Wil03}: the Lyapunov exponent is given by 
$\lambda=\langle Y\rangle/m$, with $Y$ satisfying (\ref{eq: 3.25}). 
In two dimensions, as we have already remarked, 
the term $x_2^{-1}$ is absent
from the second equation of (\ref{eq: 3.20}), and
${1\over 3}\le \Gamma\le 3$ \cite[]{Meh04}).

\section{Perturbation theory}
\label{sec: 4}
We now show how to obtain an asymptotic approximation
for the Lyapunov exponent using eqs.~(\ref{eq: 3.20}) and
(\ref{eq: 3.21}). These equations are equivalent to a 
two-dimensional Fokker-Planck 
equation (\cite{vKa92}) for a probability density $P(x_1,x_2;t')$, of the form
\begin{equation}
\label{eq: 4.1}
\partial_{t'} P=D\nabla^2P-\nabla.({\bf v}P)=\hat {\cal F} P
\ .
\end{equation}
Here the diffusion constant 
$D=1$ and the drift velocity is ${\bf v}=(v_1,v_2)$ with components
$v_1=-x_1+\epsilon (\Gamma x_2^2-x_1^2)$
and $v_2=-x_2+x_2^{-1}-2\epsilon x_1x_2$.
We write $\hat {\cal F}=\hat {\cal F}_0+\epsilon \hat {\cal F}_1$, and
seek a steady-state solution satisfying $\hat {\cal F}P=0$ by
perturbation theory in $\epsilon$. 
In order to simplify the application of perturbation theory, 
it is convenient to make a transformation so that the unperturbed
Fokker-Planck operator $\hat {\cal F}_0$ is transformed into a 
Hermitian operator. Rather than proceeding to the Hermitian 
form directly, we first map the two-dimensional Fokker-Planck 
equation to a three-dimensional equation with a rotational symmetry
(we seek a solution which is invariant under rotation). 
After making this transformation, we find that the corresponding 
Hermitian operator in three-dimensional space is the Schr\" odinger
operator of an isotropic three-dimensional harmonic oscillator.
The perturbation analysis can then be performed very easily, using
the algebra of harmonic-oscillator raising and lowering operators,
described in \cite[]{Dir30}. To shorten equations, 
we will use a variant of the Dirac
notation scheme: in summary, functions $a$, $b$ are symbolised
by vectors $\vert a)$, $\vert b)$, linear operators are
denoted by a \lq hat', e.g. $\hat{\cal A}$, and the integral
over all space of the product of two functions is denoted by 
the inner product $(a\vert b)$.

In the original form, the action of the unperturbed part of the 
Fokker-Planck operator on a function $P$ is
\begin{equation}
\label{eq: 4.2}
\hat {\cal F}_0 P=(\partial_1^2+\partial_2^2)P
+\partial_1(x_1P)+\partial_2[(x_2-x_2^{-1})P]
\ .
\end{equation}
We transform this by defining the action of $\hat{\cal F}_0'$ on
a function $P'=P/x_2$ as follows
\begin{eqnarray}
\label{eq: 4.3}
\hat{\cal F}_0'P'&=&{1\over{x_2}}\hat {\cal F}_0 P
\nonumber \\
&=&{1\over{x_2}}(\partial_1^2+\partial_2^2)(x_2P')
+{1\over{x_2}}\partial_1(x_1x_2P')+{1\over{x_2}}\partial_2[(x_2^2-1)P']
\nonumber \\
&=&\partial_1[(\partial_1+x_1)P']
+{1\over{x_2}}\partial_2[x_2(\partial_2+x_2)P']
\ .
\end{eqnarray}
We now consider $\hat{\cal F}'_0$ to be an operator acting
in three-dimensional space, with cylindrical polar coordinates
$(r,\varphi,z)$. We identify $r=x_2$, and $z=x_1$, and take $P'$
to be a function which is restricted so that it 
has cylindrical symmetry, being independent of $\varphi$.
With this interpretation, we can add differentials with respect to 
$\phi$ to the definition of $\hat {\cal F}_0'$, and write
\begin{eqnarray}
\label{eq: 4.4}
\hat {\cal F}_0'&=&{1\over r}\partial_r\bigl[r(\partial_r+r)\bigr]+{1\over{r^2}}
\partial_\varphi^2+\partial_z(\partial_z+z)
=\nabla.({\bf x}+\nabla)
\end{eqnarray}
which is the Fokker-Planck operator for isotropic diffusion
in three-dimensional space (with $D=1$), with a drift velocity
${\bf v}=-{\bf x}$. Thus we have transformed the two-dimensional
Fokker-Planck equation to a three-dimensional one with a very
simple unperturbed velocity field. It is convenient
to work with Cartesian coordinates ${\bf x} = (x,y,z)$ in the three-dimensional
space, having the usual relation to the cylindrical polar coordinates $(r,\varphi,z)$.
The Fokker-Planck equation is then
\begin{eqnarray}
\label{eq: 4.5}
\partial_{t'} P'&=&\nabla^2 P'+\nabla.[({\bf r}-\epsilon{\bf v}_1')P']
=\hat {\cal F}'P'=[\hat {\cal F}_0+\epsilon \hat {\cal F}_1]P'
\end{eqnarray}
where, in Cartesian coordinates,
${\bf v}_1'$ has components
\begin{eqnarray}
\nonumber
{v}_{11}'& =&-2xz\,,\\
\label{eq: 4.6}
v_{12}' &=&-2yz\,,\\
\nonumber
v_{13}' &=&-z^2+\Gamma (x^2+y^2)\,.
\end{eqnarray}
We now transform the Fokker-Planck $\hat {\cal F}'$ operator so that
$\hat {\cal F}_0'$ is transformed into a very simple Hermitian operator, 
by writing 
\begin{equation}
\label{eq: 4.7}
\hat {\cal H}=\exp(\Phi_0/2)\hat {\cal F}'\exp(-\Phi_0/2) 
\qquad\mbox{with}\qquad
\Phi_0={\textstyle{1\over 2}}(x^2+y^2+z^2)
\ .
\end{equation}
We find (on writing $(x,y,z)=(z_1,z_2,z_3)$)
\begin{equation}
\label{eq: 4.8}
\hat {\cal H}_0=\exp(\Phi_0/2)\hat {\cal F}_0'\exp(-\Phi_0/2)
=\sum_{j=1}^3 [\partial^2_{z_j}-{\textstyle{1\over 4}}z_j^2
+{\textstyle{1\over 2}}]
\end{equation}
so that $\hat {\cal H}_0$ is (apart from a negative
multiplicative factor) the Hamiltonian operator for a 
three-dimensional harmonic oscillator.
The spectrum of $\hat {\cal H}_0$ is the set of non-positive 
integers $(0,-1,-2,-3,..)$. The eigenfunctions of $\hat {\cal H}_0$ 
are generated by raising and lowering operators \cite[]{Dir30}:
\begin{equation}
\label{eq: 4.9}
\hat a_i={\textstyle{1\over 2}}z_i+\partial_{z_i}
\ ,\ \ \ 
\hat a_i^+={\textstyle{1\over 2}}z_i-\partial_{z_i}
\ .
\end{equation}
The transformed perturbation operator is
\begin{eqnarray}
\label{eq: 4.10}
\hat {\cal H}_1&=&-\sum_{j=1}^3 \hat a_j^+ \hat v'_{1j}
\nonumber \\
&=&2\hat a_1^+\hat z_1\hat z_3+2\hat a_2^+\hat z_2\hat z_3
-\hat a_3^+[\hat z_3^2-\Gamma(\hat z_1^2+\hat z_2^2)]
\ .
\end{eqnarray}
Instead of solving the Fokker-Planck equation $\hat {\cal F}'P'=0$
we attempt to solve $\hat {\cal H}\,Q=0$, where $Q=\exp(\Phi_0/2)P'$.
 
Now consider how to obtain the Lyapunov exponent from 
the function $Q$. We have $\lambda_1=\gamma \epsilon \langle z_3\rangle$.
We calculate the average of $z_3$ as follows
\begin{eqnarray}
\label{eq: 4.11}
\langle z_3\rangle&=&\int_0^\infty {\rm d}r\int_{-\infty}^\infty {\rm d}z\ 
P(r,z)
\nonumber \\
&=&\int_0^\infty r\,{\rm d}r\int_0^{2\pi}{\rm d}\varphi\int_{-\infty}^\infty 
{\rm d}z\ z\, P'(r,z)
\nonumber \\
&=&\int_{-\infty}^\infty \int_{-\infty}^\infty \int_{-\infty}^\infty 
{\rm d}x\ {\rm d}y\ {\rm d}z\ z\, \exp(-\Phi_0/2)Q(x,y,z)
\ .
\end{eqnarray}
Now we change the notation, using a variant of the Dirac notation 
to represent the function $Q$ by a \lq ket vector' $\vert Q)$.
Allowing for the possibility that $\vert Q)$ is not normalised,
we write
\begin{equation}
\label{eq: 4.12}
\langle z_3\rangle
={(\phi_{000}\vert \hat z_3\vert Q)\over{(\phi_{000}\vert Q)}}
={(\phi_{000}\vert \hat a_3\vert Q)\over{(\phi_{000}\vert Q)}}
\end{equation}
where $\vert \phi_{000})$ is the ground-state eigenfunction
of ${\cal H}_0$, given by the function 
$\exp(-\Phi_0/2)=\exp[-(z_1^2+z_2^2+z_3^2)/4]$.

\begin{figure}
\vspace*{4mm}
\centerline{\includegraphics[width=\textwidth,clip]{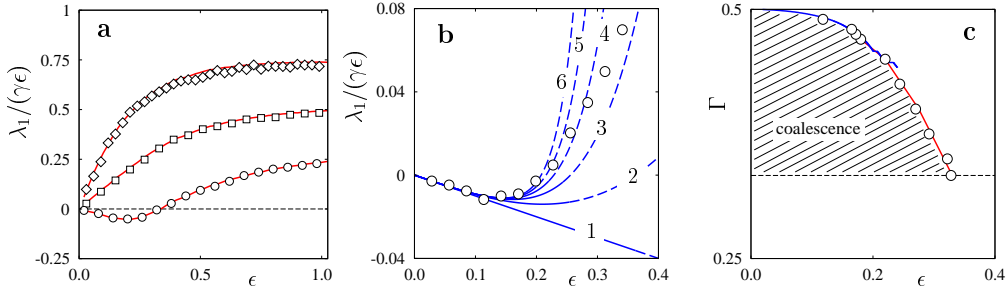}}
\caption{\label{fig: 2} 
{\bf a} Lyapunov exponent as a function
of $\epsilon$: results 
from eqs.~(\ref{eq: 3.20}), (\ref{eq: 3.21}) and (\ref{eq: 3.22})  
are shown as solid lines, those
from direct simulations as symbols, $\Gamma=1/3$ ($\circ$), $\Gamma=1$
($\Box$), and $\Gamma=2$ ($\diamond$).
{\bf b} Lyapunov exponent
$\lambda_1/(\epsilon\gamma)$ versus $\epsilon$ for $\Gamma=0.45$. 
Shown are results from simulations ($\circ$) as well as from
the asymptotic series (\ref{eq: 4.22}) summed to orders
$l_{\rm max}=1,\ldots,6$: up to $\epsilon^\ast(l_{\rm max})$ full lines
and from $\epsilon^\ast(l_{\rm max})$ dashed lines.
{\bf c} Phase diagram in the $\epsilon$-$\Gamma$-plane.
Shown are results from numerical simulations of (\ref{eq: 2.1}),
$\circ$, summation of the asymptotic series (\ref{eq: 4.22}) summed to
$k^\ast(\epsilon,\Gamma)$, blue line, as well as results
from Langevin simulations (red line).
}
\end{figure}

We calculate $\vert Q)$ by perturbation theory: writing
\begin{equation}
\label{eq: 4.13}
\vert Q)=\vert Q_0)+\epsilon \vert Q_1)+\epsilon^2\vert Q_2)+...
\end{equation}
we find that the functions $\vert Q_k)$ satisfy the recursion relation
\begin{equation}
\label{eq: 4.14}
\vert Q_{k+1})=-\hat {\cal H}_0^{-1}\hat {\cal H}_1 \vert Q_k)
\ .
\end{equation}
At first sight this appears to be ill-defined because one of the
eigenvalues of ${\cal H}_0$ is zero, so that the inverse of
${\cal H}_0$ is only defined for the subspace of functions which
are orthogonal to the ground state, $\vert \phi_{000})$.
However, because all of the components of ${\cal H}_1$ have
a creation operator as a left factor, the function $\hat{\cal H}_1\vert \psi)$
is orthogonal to $\vert \psi)$ for any function $\vert \psi )$, so
that (\ref{eq: 4.14}) is in fact well-defined.
The iteration starts with $\vert Q_0)=\vert \phi_{000})$.
The functions $\vert Q_k)$ should all have rotational symmetry
about the $z$-axis. The angular-momentum operator
$\hat {\cal J}_3=\hat p_1 \hat z_2-\hat p_2 \hat z_1$
commutes with both $\hat {\cal H}_0$ and $\hat {\cal H}_1$
(that is, $[\hat {\cal H}_0,\hat {\cal J}_3]=0$ and 
$[\hat {\cal H}_1,\hat {\cal J}_3]=0$ 
where we use square brackets for the commutator, 
$[\hat {\cal A},\hat {\cal B}]=
\hat {\cal A}\hat {\cal B}-\hat {\cal B}\hat {\cal A}$).
The operators $\hat {\cal H}_0$ and $\hat {\cal H}_1$ can therefore 
be simultaneously reduced to block diagonal form, with blocks
labelled by eigenvalues of $\hat {\cal J}_3$. We can restrict
ourselves to the subspace where the eigenvalue of $\hat {\cal J}_3$
is zero. The functions of this subspace
are generated from the ground state using transformed raising
and lowering operators, defined as follows:
\begin{eqnarray}
\label{eq: 4.15}
\hat{\alpha}_+ &=& \frac{1}{\sqrt{2}} (\hat{a}_1 + {\rm i} 
\hat{a}_2)\,,\qquad
\hat{\alpha}_- = \frac{1}{\sqrt{2}} (\hat{a}_1 - {\rm i} \hat{a}_2)
\ .
\end{eqnarray}
The transformed operators $\hat a_+$ and $\hat a_-$ satisfy
$[\hat a_\pm,\hat a^\dagger_\pm]=\hat I$ (where $\hat I$ is the 
identity operator), which is the fundamental
relation describing harmonic-oscillator raising and lowering
operators.
Expressing $\hat {\cal H}_0$ and $\hat {\cal J}_3$ in terms 
of these operators, we find
\begin{equation}
\label{eq: 4.16}
\hat{\cal J}_3 = \hat{\alpha}_-^\dagger\hat{\alpha}_- 
- \hat{\alpha}_+^\dagger \hat{\alpha}_+
\ ,\ \ \ 
\hat{\cal H}_0 =  -(\hat{\alpha}_-^\dagger\hat{\alpha}_- 
+ \hat{\alpha}_+^\dagger \hat{\alpha}_+ +
a_3^\dagger a_3)\,.
\end{equation}
Using results from \cite{Dir30}, we see that
both $\hat {\cal H}_0$ and $\hat {\cal J}_3$ are linear combinations 
of harmonic-oscillator Hamiltonians, $\hat a_-^\dagger\hat a_-$, 
$\hat a^\dagger_+\hat a_+$ and $\hat a^\dagger_3\hat a_3$.
The $n$th eigenfunction $\vert\phi_n)$ of a harmonic-oscillator Hamiltonian 
$\hat a^\dagger\hat a$ is obtained from its ground state $\vert \phi_0)$ 
by repeated application of the raising operator $\hat a^\dagger$: 
\begin{equation}
\label{eq: 4.16a}
\vert \phi_n)={1\over{\sqrt{n!}}}(\hat a^\dagger)^n\vert \phi_0)
\end{equation}
and this eigenfunction has eigenvalue $n$.
Thus eigenfunctions of $\hat{\cal H}_0$ and $\hat{\cal J}_3$ 
with zero angular momentum are constructed as follows
\begin{equation}
\label{eq: 4.17}
|\psi_{nm}) = \frac{1}{n!} \frac{1}{\sqrt{m!}} 
(\hat \alpha_-^\dagger)^n
(\hat \alpha_+^\dagger)^n
(\hat a_z^\dagger)^m |\phi_{000})
\end{equation}
for $n=0,1,\ldots$ and $m=0,1,\ldots$. The corresponding 
eigenvalues of $\hat {\cal H}_0$ are $-2n-m$.  
The functions $\vert Q_k)$ 
are expanded in terms of the $\vert \psi_{nm})$,
with coefficients $a_{nm}^{(k)}$:
\begin{equation}
\label{eq: 4.18}
\vert Q_k)=\sum_{n=0}^\infty \sum_{m=0}^\infty a^{(k)}_{nm}\vert
\psi_{nm})
\ .
\end{equation}
By projecting equation (\ref{eq: 4.14}) onto the vector
$\vert\psi_{nm})$ and using the fact that the eigenvectors
$\vert \psi_{n'm'})$ of $\hat {\cal H}_0$ form a complete
basis, the iteration 
can be expressed as follows [for $(n,m)\ne (0,0)$]:
\begin{equation}
\label{eq: 4.19}
a^{(k+1)}_{nm}
=\sum_{n'=0}^\infty\sum_{m'=0}^\infty
\frac{(\psi_{nm}\vert \hat {\cal H}_1\vert \psi_{n'm'})}{2n+m}
a^{(k)}_{n',m'}
\ .
\end{equation}
The matrix elements $(\psi_{nm}\vert\hat{\cal H}_1\vert\psi_{n'm'})$
are readily computed using the algebraic properties of the 
raising and lowering operators, as discussed in \cite{Dir30}. 
The coefficients $a^{(k)}_{nm}$ are then calculated recursively. 
This allows us to obtain the functions $|Q_k)$.
The lowest order is
$|Q_0) = |\phi_{000})$. Its contribution to $\lambda_1$
vanishes in view of (\ref{eq: 4.10}).  The leading order is
\begin{eqnarray} 
\label{eq: 4.20}
|Q_1) 
&=& -\frac{4}{3} |\psi_{11}) - |\psi_{01}) - \frac{\sqrt{6}}{3} |\psi_{03})
   +2\Gamma |\psi_{01}) + \frac{2\Gamma}{3} |\psi_{11})\,.
\end{eqnarray}
The next order, $|Q_2)$,  does not contribute to $\lambda_1$ since
$\hat{\cal H}_1|Q_1)$ does not
contain $|\psi_{01})$. In fact,
only odd orders contain $|\psi_{01})$ and thus 
give non-zero contributions
to $\lambda_1$.  We also find that the denominator 
in (\ref{eq: 4.13}) is unity at all orders. 
The final result is:
\begin{equation}
\label{eq: 4.21}
\lambda_1 =\gamma\epsilon \sum_{l=1}^\infty c_l(\Gamma)\, \epsilon^{2l-1}
\end{equation}
where the first five coefficients $c_l(\Gamma)$ are
\begin{eqnarray}
\label{eq: 4.22}
c_1(\Gamma) &=& -1+2\Gamma         \\
c_2(\Gamma) &=& -5+20\, \Gamma-16\,\Gamma^2           \nonumber\\
c_3(\Gamma) &=& -60+360\,\Gamma-568\,\Gamma^2+272\,\Gamma^3           \nonumber\\
c_4(\Gamma) &=& -1105 + 8840\,\Gamma
-61936/3\,\Gamma^2+58432/3\,\Gamma^3 -19648/3\,\Gamma^4
\nonumber\\
c_5(\Gamma)
&=&-27120+271200\,\Gamma-7507040/9\,\Gamma^2+3492160/3\,\Gamma^3\nonumber\\
&&\hspace*{5mm}-2316032/3\,\Gamma^4+ 1785856/9 \,\Gamma^5
\nonumber\\
&\vdots&\nonumber
\end{eqnarray}
The coefficients in (\ref{eq: 4.22}) have a growth which is 
typical of asymptotic series, as discussed by \cite{Din74}. 
Figure~\ref{fig: 2}{\bf b} shows
approximations to the Lyapunov exponent $\lambda_1$ for
$\Gamma = 0.45$.  Shown are six different
partial sums of the series (\ref{eq: 4.22}), including terms
up to $l_{\rm max}$, with $l_{\rm max} = 1,\ldots,6$. 
For a given value of $\epsilon$, there is an optimal
value of $l_{\rm max}$, which we term $l_{\rm max}^\ast$,
defined by the criterion that the term in (\ref{eq: 4.22}) 
with index $l_{\rm max}^\ast$
is smallest in magnitude. The function $l_{\rm max}^\ast(\epsilon)$ 
can be inverted, its inverse $\epsilon^\ast(l_{\rm max})$ 
is the value of $\epsilon$ for which the $l_{\rm max}$ term is optimal.
For values of $\epsilon$
less than the $\epsilon^\ast(l_{\rm max})$ the results
are shown as solid lines. Beyond this optimal value
of $\epsilon$, the results are shown as dashed lines.
The results show that the series agrees well with the numerical
simulation up $\epsilon^\ast$, as would be expected for an
asymptotic series.

%One notable difference between the three-dimensional calculation
%presented here and the two-dimensional case considered in 
%\cite[]{Meh04} is that in the two-dimensional case the 
%phase boundary has a power series in $\epsilon$ which vanishes
%identically (and the phase line is therefore non-analytic). 

The phase boundary in the $\epsilon$-$\Gamma$-plane is determined
by the condition $\lambda_1= 0$. Figure~\ref{fig: 2}{\bf c} shows
this phase boundary as determined from truncating
the series (\ref{eq: 4.22}) at the optimal order
$l_{\rm max}^\ast(\epsilon)$ (blue line). This asymptotic
result is shown for values of $\epsilon$ up to $\approx 0.2$.
Beyond this range, the asymptotic approximation becomes
increasingly inaccurate. Also shown, in the same plot,
are results obtained from the Langevin equations 
[eqs.~(\ref{eq: 3.20}), (\ref{eq: 3.21}) and (\ref{eq: 3.22})],
and from direct  simulations ($\circ$). The results show that the
coalescing phase disappears as the effect of inertia is 
increased: the coalescing disappears altogether
for the case of pure potential flow ($\Gamma={1\over 3}$) at 
$\epsilon\approx 0.33$.

One notable difference between the three-dimensional calculation
presented here and the two-dimensional case considered in 
\cite[]{Meh04} is that in the two-dimensional case the 
phase boundary has a power series in $\epsilon$ which vanishes
identically (and the phase line is therefore non-analytic).

\section{Effect of dispersion of particle masses}
\label{sec: 5}

In most naturally occurring aerosols the suspended particles
have different mass $m$, and particles of differing sizes
will also have different values of the damping coefficient
$\gamma$. It is important to consider whether particles
still have a tendency to coalesce even when the particles have
differing values of $m$ and $\gamma$: we argue that
the path coalescence effect is not destroyed by small
mass dispersion. The argument can be adapted to
dispersion of the damping coefficient, reaching the same
conclusion.

Assume that the path-coalescence effect occurs for particles of
mass $m$. Compare the motion of this reference particle with that
of an initially nearby particle with mass $m+\delta m$.
The reference particle has equation of motion
\begin{equation}
\label{eq: 5.1}
m\ddot x=-\gamma m\dot x+f(x,t)
\ .
\end{equation}
Writing $f(x+\delta x,t)=f(x,t)+F(t)\delta x+O(\delta x^2)$, the
equation of motion for the other particle is 
\begin{equation}
\label{eq: 5.2}
(m+\delta m)(\ddot x+\delta \ddot x)
=-\gamma m(\dot x +\delta \dot x)+f(x,t)+F(t)\delta x +O(\delta x^2)
\ .
\end{equation}
Collecting the terms which are first order in $\delta x$, we 
obtain a linearised equation of motion for $\delta x$:
\begin{equation}
\label{eq: 5.3}
-m\delta \ddot x-\gamma m\delta \dot x+F(t)\delta x
=\delta m\, \ddot x
\ .
\end{equation}
This is an inhomogeneous differential equation for the separation
$\delta x$ between two particles, with a driving term proportional
to their mass difference $\delta m$. 
The solution of this equation can be constructed from a Green's
function satisfying $G(t,t_0)$
\begin{equation}
\label{eq: 5.4}
-m{{\rm d}^2G\over{{\rm d}t^2}}
-\gamma m{{\rm d}G\over{{\rm d} t}}+F(t)G=\delta(t-t_0)
\end{equation}
with $G(t,t_0)=0$ for $t<t_0$. The solution of equation (\ref{eq: 5.3}) is
\begin{equation}
\label{eq: 5.5}
\delta x(t)=\delta m\int_0^t{\rm d}t'\ G(t,t')
{{\rm d}^2 x(t')\over{{\rm d}t'^2}}
\ .
\end{equation}
For $t>t_0$, equation (\ref{eq: 5.4}) is the equation for small displacements
of trajectories of particles with the same mass. We know that in 
the path-coalescing phase the solutions have a negative value of the 
Lyapunov exponent $\lambda_1$, and that they therefore decay
exponentially at large time. 
In the case where $G(t,t')$ is bounded by an exponentially
decreasing function, such that 
$\vert G(t,t')\vert <A\exp(-\lambda_1\vert t-t'\vert)$, 
equation (\ref{eq: 5.5}) remains finite as $t\to \infty$.
For sufficiently large $A$, the probability of this inequality
being violated is extremely small.
This indicates that in the path-coalescing
phase the solution (\ref{eq: 5.5}) remains finite as $t\to \infty$,
except for very rare events. The conclusion is that, when
$\lambda_1 <0$, two initially close particles with nearly
equal mass will remain in close proximity for a very
long time.

\section{Discussion}
\label{sec: 6}

In this paper we described the path-coalescence transition,
and showed that the transition point is determined by the change 
of sign of a Lyapunov exponent. We showed that in general the Lyapunov
exponent is determined from an expectation value of a 
variable of a simple dynamical system, equations (\ref{eq: 2.12}), 
which is driven by stochastic forcing functions. 
We considered the solution of these equations in a particular
limiting case, where the dimensionless parameters 
satisfy $\chi \ll \nu \ll 1$,  by mapping the 
continuous differential equations into a pair of coupled Langevin 
equations. We used these Langevin equations to produce a rather
complete description of the phase transition in that limit. 
The remainder of these concluding remarks
discuss how equations (\ref{eq: 2.12}) can be used in the
case where these inequalities are not satisfied.

In order to solve these differential equations it is necessary
to characterise the statistics of the stochastic driving terms
$F'_{ij}(t)$. These terms contain information about the strain-rate
of the field evaluated at a point along the reference particle
trajectory. There are two possibilities:

\begin{description}

\item Case A: the statistics of the 
strain-rate tensor along a trajectory may be indistinguishable 
from those sampled along a randomly chosen trajectory.

\item Case B: the 
trajectory may select regions where the strain-rate tensor
has atypical properties, for example by tracking points where
the velocity vector ${\bf u}$ vanishes.

\end{description}

If case A is realised there are two further possibilities:

\begin{description}

\item Case A1: the trajectory ${\bf r}(t)$ is sufficiently
slowly moving that the displacement over time $\tau$ is small
compared to $\xi$. In this case the statistics of $F'_{ij}$
are those of a randomly chosen static point, and the correlation
time of $F'_{ij}(t)$ will be $\tau$.

\item Case A2: if the trajectory ${\bf r}(t)$ is moving sufficiently
rapidly that its displacement in time $\tau $ is large compared
to $\xi$, then the correlation time of $F'_{ij}(t)$ will be smaller
than $\tau$ because the loss of correlations results primarily
from changing the position at which $\partial u_i/\partial x_j({\bf r},t)$
is sampled.

\end{description}

The limit which was investigated in detail in this paper 
($\chi\ll\nu\ll 1$) is an example of case A1. 
In cases where $\chi$ and $\nu$ approach different limits however, 
all three possibilities can occur in the system described by
equations (\ref{eq: 1.1}) to (\ref{eq: 1.3}).
%, making this a very rich model.

\bibliographystyle{jfm}

\begin{thebibliography}{12}
\bibitem[Falkovitch, {\sl et. al} (2001)]{Fal03}
{\sc Balkovsky, E., Falkovitch, G. \& Fouxon, A.}, 2003,
{Intermittent distribution of inertial particles in turbulent flows},
{\it Phys. Rev. Lett.}, {\bf 86}, 2790-3.


\bibitem[Baxendale \& Harris (1986)]{Bax86}
{\sc Baxendale, P. \&  Harris, T.}, 1986, 
{Isotropic stochastic flows}, 
{\it Ann. Probab.}, {\bf 14}, 1155-79.


\bibitem[Cressman {\sl et al.} (2004)]{Cre04}
{\sc Cressman, J. R., Goldburg, W. I. \& Schuhmacher, J.}, 2004, 
{Dispersion of tracer particles in a compressible flow},
{\it Europhys. Lett. } {\bf 66}, {219}.


\bibitem[Deutsch (1985)]{Deu85}
{\sc Deutsch, J. M.}, 1985,
{Aggregation-disorder transitio induced by fluctuating random forces},
{\it J. Phys. A}, {\bf 18}, 1457.


\bibitem[Dingle (1973)]{Din74}
{\sc Dingle, R. B.}, 1974,
{Asymptotic expansions:  their derivation and interpretation},
Academic Press, New York.


\bibitem[Dirac (1930)]{Dir30}
{\sc Dirac, P. A. M.}, 1930,
{The principles of quantum mechanics},
Oxford University Press.


\bibitem[Eckmann \&  Ruelle (1979)]{Eck79}
{\sc Eckmann, J-P. \& Ruelle, D.}, 1985,
{\it Rev. Mod. Phys.}, {\bf 57}, 617-56.


\bibitem[Elperin {\sl et al}(1996)]{Elp96}
{\sc Elperin, T., Kleeorin, N. \& Rogachevskii, I.}, 1996, 
{Self-excitation of fluctuation of inertial particle concentration
in turbulent fluid flow},
{\it Phys. Rev. Lett.}, {\bf 77}, 5373.


\bibitem[Fessler {\sl et al}(1994)]{Fes94} 
{\sc Fessler, J. R., Kulick, J. D. \&  Eaton, J. K.}, 1994,
{Preferential concentration of heavy particles in a 
turbulent channel flow}, 
{\it Phys. Fluid}, {\bf 6}, {3742}.


\bibitem[Frisch (1997)]{Fri97}
{\sc Frisch, U.}, 1997,
{Turbulence},
Cambridge Univeristy Press.


\bibitem[Klyatskin \& Gurarie (1999)]{Kly99} 
{\sc Klyatskin, V. I. \& Gurarie, D.}, 1999,
{Coherent phenomena in stochastic dynamical systems},
{\it Physics Uspekhi}, {\bf 42}, 165.


\bibitem[Landau \& Lifshitz (1958)]{Lan58}
{\sc Landau, L. D. \&  Lifshitz, E. M.}, 1958,
{Fluid Mechanics}, Pergamon, Oxford.


\bibitem[Le Jan (1985)]{LeJ85}
{\sc Le Jan, Y.}, 1985,
{On isotropic Brownian motions},
{\it Z. Wahsch. verw. Gebiete}, {\bf 70}, 609-20.


\bibitem[Maxey (1987)]{Max87}
{\sc Maxey, M. R.}, 1987,
{The gravitational settling of aerosol particles in homogenous
turbulence and random flow fields}
{\it J. Fluid Mech.} {\bf 174}, 441-465.


\bibitem[Mehlig \& Wilkinson (2004)]{Meh04}
{\sc Mehlig, B. \& Wilkinson M.}, 2004,
{Coagulation by random velocity fields as a Kramers problem},
{\it Phys. Rev. Lett.}, {\bf 92}, 250602.


\bibitem[Piterbarg (2001)]{Pit01}
{\sc Piterbarg, L.}, 2001, 
{The top Lyapunov exponent for a stochastic flow modelling 
the upper ocean turbulence}, 
{\it SIAM J. Appl. Math.}, {\bf 62}, 777-800.


\bibitem[Shaw (2003)]{Sha03}
{\sc Shaw, R. A.}, 2003,
{Particle-turbulence  interactions in atmospheric clouds}, 
{\it Annu. Rev. Fluid Mech.}, {\bf 35}, 183-227.


\bibitem[Sigurgeirsson \& Stuart (2002)]{Sig02}
{\sc Sigurgeirsson, H. and Stuart, A. M.}, 2002,
{A model for preferential concentration}, 
{\it Phys. Fluids}, {\bf 14}, 4352-61.


\bibitem[Sommerer \& Ott (1993)]{Som93}
{\sc Sommerrer, J \&  Ott, E.}, 1993, 
{Particles floating on a random flow: a dynamically comprehensible 
physical fractal}, 
{\it Science}, {\bf 359}, {334}.


\bibitem[Wilkinson \& Mehlig (2003)]{Wil03}
{\sc Wilkinson, M. \& Mehlig, B.}, 2003,
{Path-coalescence transition and its applications},
{\it Phys. Rev. E}, {\bf 68},  040101(R).


\bibitem[van Haarlem {\sl et al.} (1998)]{vHa98}
{\sc van Haarlem, B., Boersma, B. J. \&  Nieuwstadt, F. T. M.}
{Direct numerical simulations of particle deposition onto
a free-slip and no-slip surface}
{\it Phys. Fluids}, {\bf 10}, 2608-2620.


\bibitem[van Kampen (1992)]{vKa92}
{\sc van Kampen, N. G.}, 1992,
{Stochastic processes in physics and chemistry},
North-Holland.


\end{thebibliography}

\end{document}